\documentclass{elsart}
\usepackage[dvips]{graphicx}

\begin{document}
\begin{flushright} 
SLAC-PUB-9111 \\
Jan 7 2002
\end{flushright}
\vspace{3cm}
\begin{frontmatter}
\title{A Monitor of the Focusing Strength of Plasma Lenses using MeV Synchrotron
Radiation\thanksref{DOE}}
\thanks[DOE]{Work supported by the U.S. Department of Energy, Contract DE-AC03-
76SF00515}

\author{Clive~Field\corauthref{field}},
\corauth[field]{Corresponding author. Tel.: + 1-650-926-2694; fax: +1-650-926-4178}
\ead{sargon@slac.stanford.edu}
\author{Gholam~Mazaheri} and
\author{Johnny~S.T.~Ng}

\address{Stanford Linear Accelerator Center, Stanford University, Stanford, CA 94309, U.S.A.}
\begin{abstract}
The focusing strength of plasma lenses used with high energy electron or positron beams can
give rise to synchrotron radiation with critical energies in the MeV range. A method is described
for measuring the characteristic energy of this radiation as a way of monitoring the strength of the
focus. The principle has been implemented in a  plasma lens experiment with a 28.5 GeV
positron beam.
\end{abstract}

\end{frontmatter}

\begin{center}
Submitted to Nuclear Instruments and Methods in Physics Research A
\date{today}
\end{center}
\clearpage
\section{Introduction}

     An experiment has been carried out at SLAC to study the focusing of electron and
positron beams by thin targets of plasma \cite{JNgetal}. Focusing systems of this type may be
used to boost luminosity at future linear colliders by sharpening their final focus. If so, a means
of monitoring the operation of the lenses must be developed, at least during single beam tuning.
As part of the experimental study, a monitor based on the synchrotron radiation from the plasma
lens has been developed. It was used both online, as a tuning tool, and  
in data analysis to
validate the measurements of focusing by wire scanning of the beam spots.

     The SLAC Final Focus Test Beam (FFTB) \cite{FFTB}, where the experiment was done,
is a model of the final focus of a linear collider and is presently operated up to 28.5 GeV. With
these energies, plasma lenses are strong enough that their synchrotron radiation exhibits critical
energies, ${\rm E_c}$, in the MeV range. The radiation is therefore quite penetrating, and the
depth of penetration was used as a measure of ${\rm E_c}$.

     Above the absorption edges, the gamma ray absorption length increases with 
increasing
energy until the onset of pair production which reverses this trend. Common particle calorimeter
materials, like Pb, have high atomic number, Z, and so their pair production cross section rises
quickly above threshold, becoming dominant above a few MeV. For these materials, the slope of
the graph of penetration depth against energy quickly changes sign and higher energy gamma
rays tend to interact at shallower depths.

\begin{figure}
\begin{center}
\includegraphics*[width=8cm]{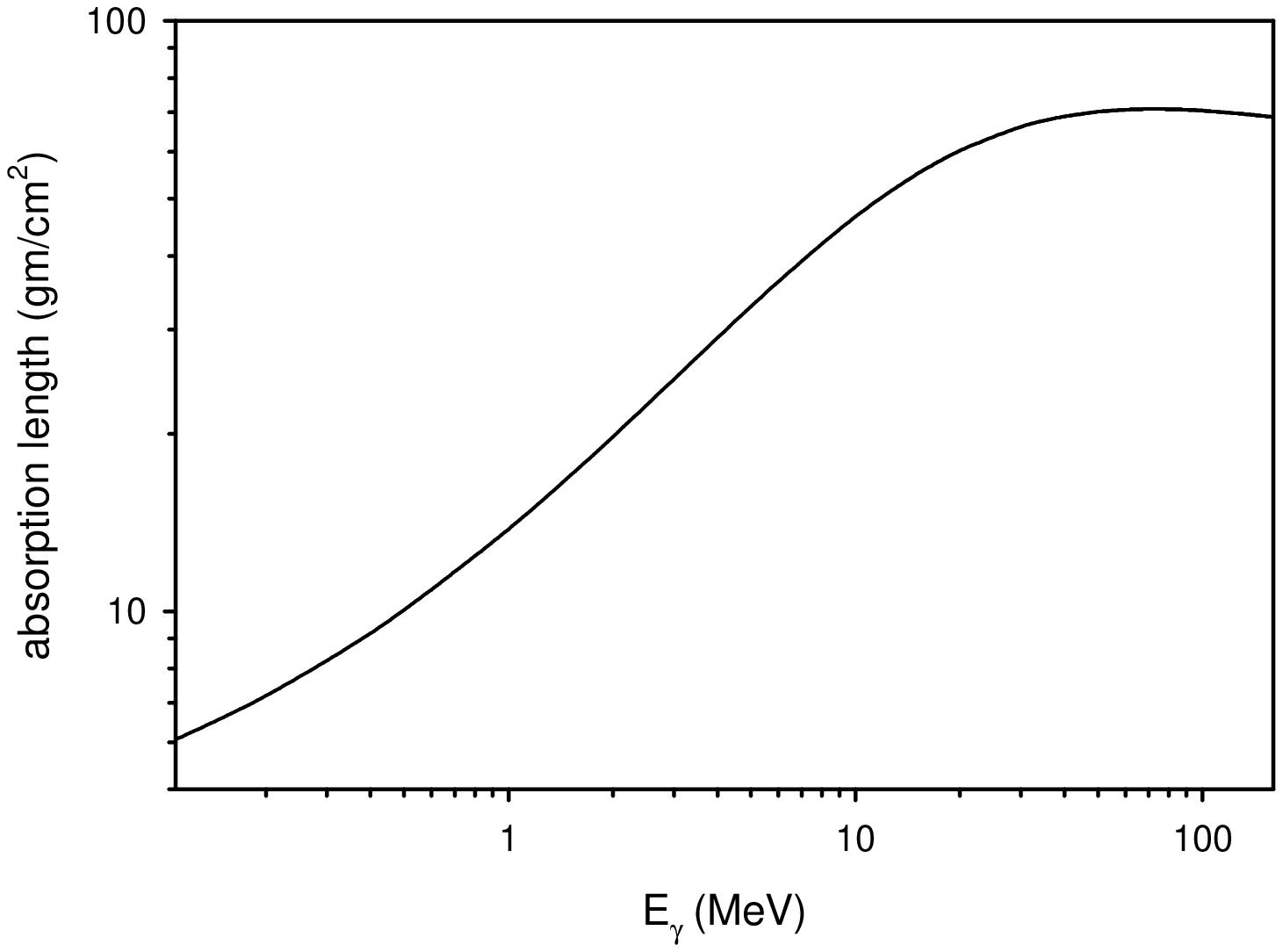}
\end{center}
\caption{Gamma ray absorption length in the range 100 keV to 
150 MeV for polyethylene.}
\label{absorption-length}
\end{figure}

     However, the synchrotron radiation spectrum has a significant component up 
to $\sim 5 \times{\rm E_c}$. To accommodate ${\rm E_c}$ in the MeV range
one must use materials of
low Z, such as carbon, where the dominance of pair production is postponed until energies of 50
MeV or above are reached \cite{XCOM}. This is illustrated in Fig.~\ref{absorption-length}.

     A calibration beam of synchrotron radiation with a critical energy in the range of interest 
--- or even a suitable monochromatic photon beam of tunable energy in this range --- is rarely
available, and so one must rely on particle transport codes such as EGS4 \cite{EGS4} to model
the detector response. At critical energies above $\sim$50 MeV, as could be the case with a
future linear collider, these codes  allow the technique to be extended into the range where the
shower profile from the synchrotron radiation lengthens logarithmically with increasing energy,
and low-Z material would not be advantageous.   

     The study of plasma lenses at presently available energies is hampered by the physical
limits of practical beam-spot size measuring techniques. Wire scanning with carbon fibers has
been the tool of choice, but plasma focusing can easily produce spots that instantaneously
vaporize the fiber \cite{JNgetal,wscanner}. Monitoring of the focusing strength by the use of
synchrotron radiation will permit a substantial extension of the work on plasma lenses.
 
\section{Experimental Setup}

     The experiment tested the focusing of electron and positron beams at 28.5 GeV, produced
by the SLAC linac and delivered to the FFTB at ${\rm1.5 \times10^{10}}$ particles per pulse,
with pulse duration, typical of linear colliders, of 4 ps FWHM. The beam was intercepted by a 3
mm thick jet of nitrogen whose density, for the data reported here, was set to correspond to 0.25
Atm. The resulting ionization formed a plasma that focused the beam with an effective focal
length in the range of millimeters \cite{JNgetal}, accompanied by the emission of a burst of
synchrotron radiation.

     The beam dimensions downstream of the jet were measured in a standard procedure
\cite{wscanner} by a carbon fiber wire scanner. The procedure was to use magnetic dipoles to
scan the beam, in micron steps, across a carbon fiber. The resulting Bremsstrahlung continued
down the beam line along with the charged beam. Eventually the charged beam was deflected to
its dump, and the gammas were allowed to leave the vacuum pipe, 30 meters from the gas jet,
through a thin window. There they entered the front of a stack of absorber material in which were
embedded a set of detectors. The separation between charged and neutral beams was 26 cm at
this point. The final detector, after 4 radiation lengths of material, was an air Cherenkov counter
\cite{wscanner}, and its signal, proportional to the overlap of the beam and the fiber, was used as
the wire scanner signal.

     In addition to the Bremsstrahlung, there was the MeV synchrotron radiation from the
plasma lens, and some synchrotron radiation, in the $\sim$100 keV range, from conventional
beam line elements. The material before the air Cherenkov counter absorbed this radiation, and
each of the detectors interspersed through the material was sensitive to the surviving flux at its
own depth.

\begin{figure}
\begin{center}
\includegraphics*[width=8cm]{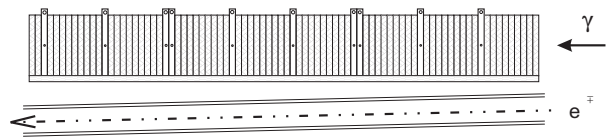}
\end{center}
\caption{Elevation view of the detector set up. The charged beam line is 
indicated descending from right to left below the detector. The ion 
chambers can be seen spaced between blocks of
polyethylene. At layers 3 and 6, a second ion chamber is shown. These 
provided independent
signals to the accelerator control system. The absorber blocks are 25 cm 
high, and the stack is 213 cm long.}
\label{elev-view}
\end{figure}

     As a convenient absorber material we chose polyethylene, n(CH$_2$), although water would
have similar absorption properties. It was available in 2.5 cm thick plates. We had these cut into
25 cm squares which we stacked, along the beam direction, in blocks 22.5 cm thick.
Fig.~\ref{elev-view} illustrates the long profile of the stack. Between each block was a planar
ion chamber. In total there were 8 layers, each with one absorber block and its ion chamber. Two
additional chambers were inserted to allow the accelerator data collection system to monitor
conditions independently. The complete stack corresponded to approximately 4 radiation lengths.

     Ion chambers were chosen as the sensitive elements primarily because of the intensity of
the signal. The electron density from the converted gamma rays could reach $2 \times 10^7~{\rm
cm^{-2}}$ per pulse, and radiation doses in the range of tens of Megarads were accumulated
during the experiment. Ion chambers held out the promise of rugged performance, while
requiring minimal intervention, and have excellent linearity. Preliminary tests showed pulse
linearity, with a nitrogen filling, above 2 nanocoulombs ${\rm cm^{-2}}$, beyond what would
be needed for the plasma focusing experiment.

     The ion chambers were of a simple double-gap window-frame construction. The bodies
were constructed of three acrylic plates 3.2 mm thick, spaced by acrylic "window-frames" 6.4
mm thick and 19 mm wide, to form two 6.4 mm collecting gaps. On the internal surfaces of the
plates was bonded a laminate of 8~$\mu{\rm m}$ aluminum on 75~$\mu{\rm m}$ Mylar, the
aluminum forming the electrodes of the chambers. Polyimide tape covered the edge of the
aluminum, extending 6 mm in from the acrylic frame. This increased the surface path length
between opposite electrodes, and so reduced the likelihood of surface discharges across the frame
material. The area remaining active in each gap was then 20 cm square. Negative high voltage
was applied to the outer plates, through a 1 megohm resistor for security. Signals from collected
electrons were picked up from the electrodes on either side of the middle plate and taken to a
preamplifier in a shielded box at the side of the chamber. The preamplifier, with a gain of 11,
was based on a KH300 chip \cite{amp}.

     The gas chosen was nitrogen, and this was fed in parallel to the two gaps at a rate of a few
cc per minute, the gas lines penetrating a Faraday enclosure around both the chamber and its
preamplifier. The preamplifier box was also in the Faraday shield, and its signals emerged by
way of a BNC connector to be carried to standard 11-bit CAMAC ADCs over about 75 m of
cable, mostly RG 214.

     Prototype tests indicated that the chambers could operate at 1000V with a collection
efficiency that was independent of the applied high voltage, and so all were operated at the same
setting, and minimal monitoring of the H.V. was needed. At this voltage, the signal at the ADC
was observed to reach peak amplitude at 50 ns, and to fall within 10\% of peak at about 360 ns.

\section{Operation and Results}

     The carbon fiber scanning of the beam spot profile was carried out at various positions
along the beam line (z coordinate) close to the focal point of the beam. The intention was to map
the convergence of the beam to the waist, and its divergence beyond there. The rms spot widths
were in the range 1.5 to 15 $\mu{\rm m}$. At any given z position, the beam profile was scanned
in about 50 steps, of which perhaps one third would be in the peak. Although the beam was
present at 10 Hz, the vacuum pumping system would permit the jet of gas to be injected only at 2
Hz. Consequently, at each scan step, data were taken for four beam pulses without any gas
injection. This provided a baseline measurement of the undisturbed beam. On the fifth pulse, the
gas, and its plasma focusing effect, were present, and the records from this pulse could be
compared with those of the previous four. Since wire scanning is not a fast procedure, this
pattern, repeated for every step of the scan, allowed us to minimize the effects of drifts of the
beam parameters.

\begin{figure}
\begin{center}
\includegraphics*[width=8cm]{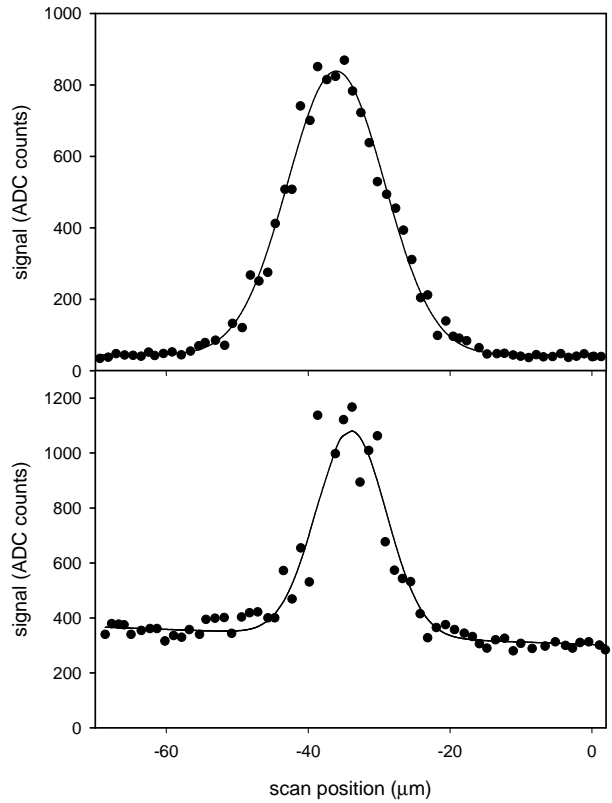}
\end{center}
\caption{Example of the signals from an ion chamber during a scan 
of the beam across a carbon fiber. In the lower plot
the plasma was present; in the upper one there was no gas injection. The 
indicated beam widths
are $5.0\pm 0.1$ and $6.9\pm 0.07 \mu{\rm m}$ respectively. Note that the 
plasma-on base line is high because of synchrotron radiation from plasma 
focusing and Bremsstrahlung from the gas.}
\label{scan}
\end{figure}

     An example of the profiles obtained from an ion chamber during a scan is shown in
Fig.~\ref{scan}. The upper graph is in the absence of nitrogen (four pulses are averaged per step).
The peak is, of course, from the beam-fiber overlap, and the lack of smoothness is caused by
fluctuations of the beam parameters. Details of the shape are strongly consistent at different
depths in the polyethylene. The fitted peak amplitude can be used as a relative measurement of
the Bremsstrahlung shower intensity at each chamber. The lower graph is from the same scan,
but from pulses with the gas jet firing. The peak again is from the carbon fiber, although with
only a single pulse per point, but in this plot the baseline is considerably higher. It has
contributions from the beam background (also visible in the baseline of the gas-off curve), from
Bremsstrahlung from the nitrogen, and from the plasma-focus synchrotron radiation. The analysis
task is to separate these.

     The beam background is easily removed by subtracting the baseline measured in the
gas-off plot. For the first ion chamber, the ``beam background'' is dominated by the low energy
synchrotron radiation from the conventional beam line dipoles. In fact, its ADC would saturate
when strong plasma focusing was added to this. For this reason, its data are discarded from the
procedure discussed below.

     The Bremsstrahlung depth profile can be obtained from the ion chamber signals
corresponding to the carbon fiber peak amplitudes. The shapes of the Bremsstrahlung depth
profiles are consistent from run to run, within 0.4\% in layer 2, and 0.2\% for deeper layers.
Beam intensity and spot size variations control the overall amplitude, but do not affect the
profile.

\begin{figure}
\begin{center}
\includegraphics*[width=8cm]{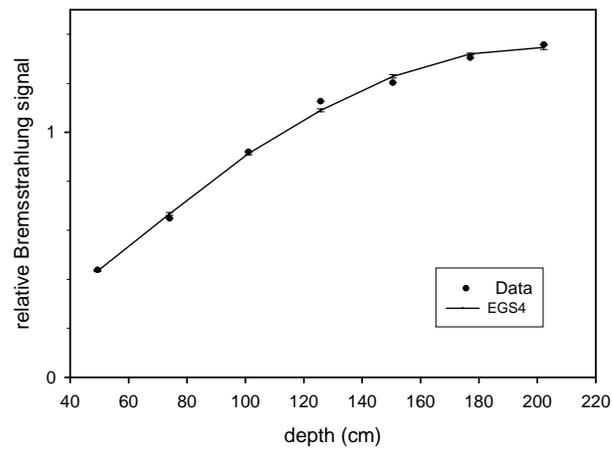}
\end{center}
\caption{Depth profile of Bremsstrahlung from the carbon fiber. Data from the ion chambers is
compared with the simulation results (connected by lines).}
\label{Brem-v-depth}
\end{figure}

     To validate the procedure, an EGS4 simulation was made of a Bremsstrahlung
spectrum incident on a model of the detector stack. A comparison is
shown (Fig.~\ref{Brem-v-depth}) of this simulation against the results
from wire scans. The average of three scans is used,
and the ordinates are simply scaled to equate the average signal strength in simulated and real
chambers --- no fit is made. The discrepancies are taken to be estimates of the relative
sensitivities of the chambers and electronics channels, with small contributions from
nonuniformity in the polyethylene. They are used to obtain correction factors for each layer. An
estimate of the uncertainties in the plotted points is made from variations between scans, and
among different EGS4 runs.

     In order to address the performance of the system for synchrotron radiation, the EGS4
code was run on the same model of the apparatus, but with a set of individual energies from 200
keV to 150 MeV. The output for each energy and layer was incorporated in a code which
simulated a dipole synchrotron radiation spectrum. It interpolated the EGS4 results at each
spectrum point. Results from critical energies in the range 0.5 to 10 MeV were examined.

\begin{figure}
\begin{center}
\includegraphics*[width=8cm]{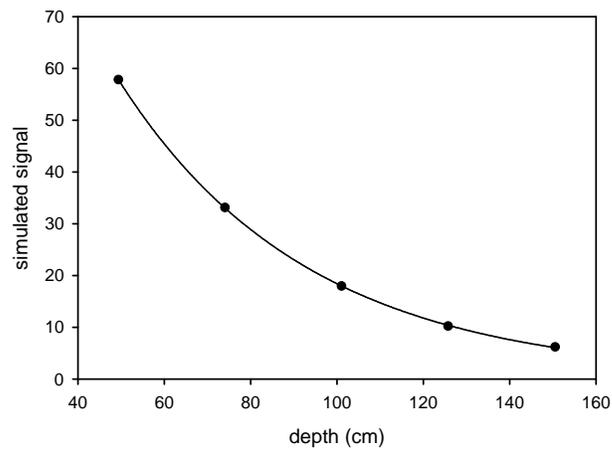}
\end{center}
\caption{Example of a simulated depth profile for synchrotron radiation. The critical energy is
4.5 MeV. The fit line is a 3-parameter exponential absorption curve.}
\label{SRsim4.5}
\end{figure}

\begin{figure}
\begin{center}
\includegraphics*[width=8cm]{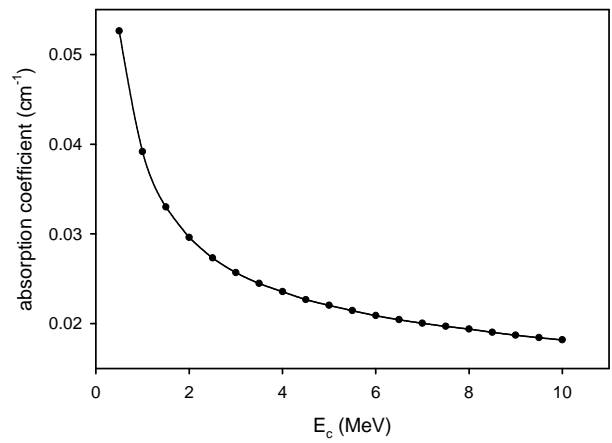}
\end{center}
\caption{Locus of effective absorption coefficient (in ${\rm cm^{-1}}$)
against critical energy, from simulation.}
\label{calib-curve}
\end{figure}

     In this experiment the final two layers saw very little signal from synchrotron radiation.
They were dominated by Bremsstrahlung, and this fact allows for a straightforward subtraction of
the Bremsstrahlung contribution to the other chambers. A simulated depth profile of the five
relevant layers is shown in Fig.~\ref{SRsim4.5}. Superimposed is the fit of a simple three
parameter exponential absorption function $y=y_0+a \exp(-bx)$, which is evidently quite
adequate. The absorption coefficient, $b$, is obviously the inverse of the effective absorption
length. The constant term, $y_0$, is retained empirically since it improves the fits. However, 
it is
always found to be very small, positive, and just significant statistically. A plot of the 
values of the absorption coefficient obtained from fits to the simulations is shown in 
Fig.~\ref{calib-curve} against the
critical energies of the simulated spectra. This is effectively a calibration curve.

\begin{figure}
\begin{center}
\includegraphics*[width=8cm]{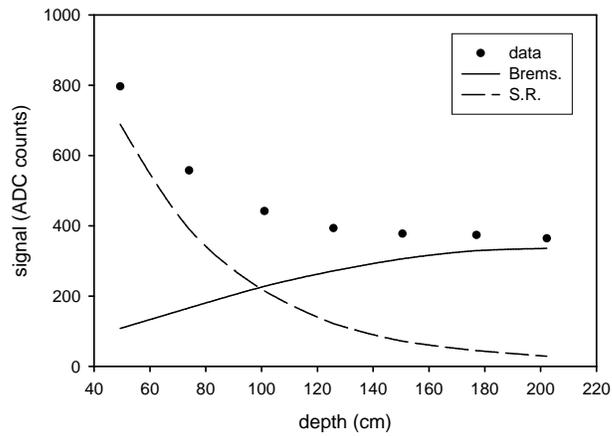}
\end{center}
\caption{Signal depth profile from a scan, after gas-off base line subtraction. The contributions
of plasma lens synchrotron radiation (dominant at shallow depths), and Bremsstrahlung
(dominant to the right of the plot), are shown as lines.}
\label{depth-profile-sr-brem}
\end{figure}

     The experimental depth profile obtained from the scan baseline (caused by
Bremsstrahlung from the gas plus plasma synchrotron radiation),
is illustrated in Fig.~\ref{depth-profile-sr-brem}. The gas-off
background has already been subtracted and the relative chamber
sensitivities corrected for. The lines in the figure are discussed below.

     The Bremsstrahlung part is now to be removed. For this purpose we note that, for critical
energies in the range 1 to 10 MeV, the ratio of synchrotron radiation in layers 8 and 7 should be
in the range 0.66 to 0.72 (from simulation). For Bremsstrahlung the layer 8 / layer 7 ratio should
be 1.027. Equating the observed signals in each of these layers to the sum of the synchrotron
radiation and Bremsstrahlung contributions, and coupling these equations with the ratios just
quoted, one can simply solve for the intensities of the two components. We obtain upper and
lower limits to the Bremsstrahlung intensity corresponding to the full 0.66 to 0.72 range of the
synchrotron radiation ratio. With the Bremsstrahlung intensity determined for the last two layers,
its full depth profile can then be subtracted from the observed signals. The separate contributions
from Bremsstrahlung and plasma-lens synchrotron radiation are indicated as lines in
Fig.~\ref{depth-profile-sr-brem}.

\begin{figure}
\begin{center}
\includegraphics*[width=8cm]{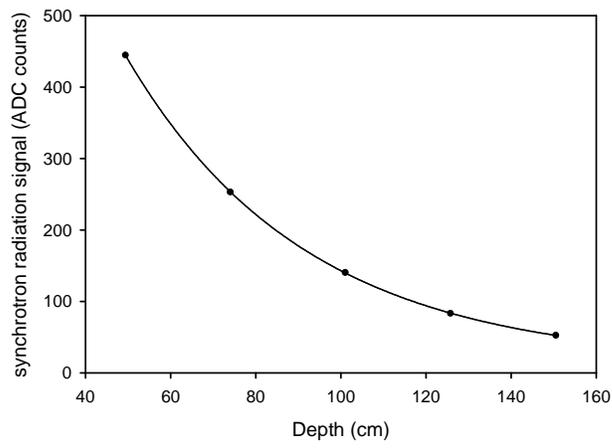}
\end{center}
\caption{Example of a synchrotron radiation depth profile derived from the base line of a scan,
with its 3-parameter exponential absorption fit.}
\label{SR-data-prof}
\end{figure}

\begin{figure}
\begin{center}
\includegraphics*[width=8cm]{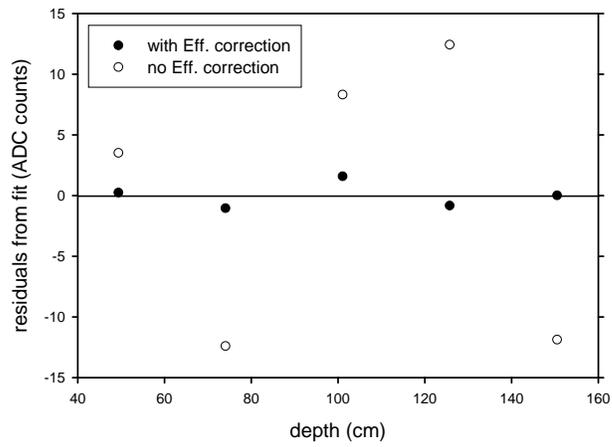}
\end{center}
\caption{Residuals from a fit, compared with a similar treatment but without applying
the layer-by-layer efficiency corrections. The corrections greatly improve the residuals.}
\label{resids}
\end{figure}

     Plasma synchrotron radiation results from a scan are illustrated
in Fig.~\ref{SR-data-prof}. The exponential absorption fit is indicated. The effect of the 
relative sensitivity corrections
of the layers is illustrated by comparing the residuals from the exponential fits, with and without
the corrections, in Fig.~\ref{resids}. It is evident that the corrections, made using the
Bremsstrahlung depth profile, substantially improve the fit to the synchrotron radiation 
profile. The approximation of using a dipole synchrotron radiation spectrum to parametrize the 
radiation from the lens is also seen to be good.

\begin{figure}
\begin{center}
\includegraphics*[width=8cm]{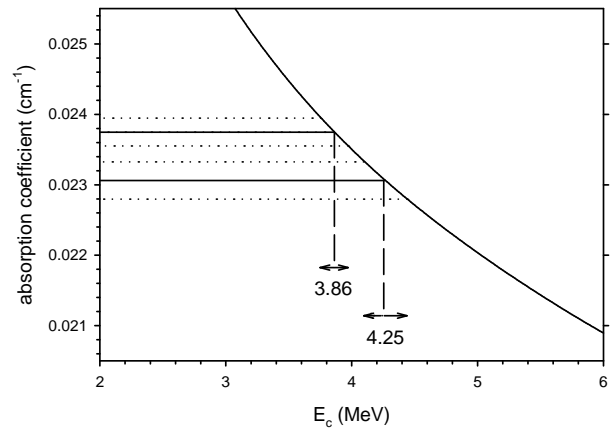}
\end{center}
\caption{Results from the exponential fits to depth profiles of two scans, one taken with the
plasma pre-excited by a laser. The fit values for the absorption coefficients are drawn on part of
Fig.~\ref{calib-curve}, the calibration curve. The ${\rm E_c}$ values read from the graph are
$3.86\pm 0.13$ and $4.25\pm 0.20$ MeV. The higher critical energy is for data where the
plasma gas was pre-excited, and the focusing was stronger.}
\label{results}
\end{figure}

     Evaluation of ${\rm E_c}$ for two scans is illustrated graphically
in Fig.~\ref{results}.
For one of the scans, the plasma was pre-excited by a laser, and the focusing was somewhat
stronger in that case. The curve is a magnified part of the calibration curve.
The synchrotron radiation absorption coefficients were determined from the fits to the
data as above, and are drawn as horizontal lines intersecting the calibration curve. The
uncertainties obtained from the fits and from the upper and lower limits of the Bremsstrahlung
subtraction are also indicated. One can read off the equivalent dipole critical energies of the two
cases: $3.86 \pm 0.13$ MeV for the less strongly focused case, and $4.25 \pm 0.20$ MeV for the
pre-excited gas. The respective absorption lengths are 42.1 cm and 43.4 cm polyethylene
equivalent. In Ref. \cite{JNgetal}, it is concluded that ${\rm E_c}$ values in this range are
consistent with the focusing strengths, 0.7 T/$\mu$m in the x plane and 4 T/$\mu$m in y,  
estimated from the beam waist profile measurement.

\section{Conclusions}

     An initial demonstration has been made of a monitoring technique for the very strong
focusing of high energy particle beams made possible by plasma lenses. The technique is not
difficult or expensive, is robust, and could be modified for a range of applications. It makes use
of the synchrotron radiation emitted by the lens, and should work over a 
range $1<{\rm E_c}<50$ MeV as implemented in this example. At higher energies, the logarithmic 
lengthening of the
electromagnetic shower with gamma ray energy would be used to monitor the profile. The
principal requirement is simply that the synchrotron radiation flux be separated sufficiently from
the charged beam that it becomes accessible.

\section{Acknowledgments}

     We thank the SLAC Plasma Lens Collaboration for their efforts in running the
experiment. We acknowledge especially the work of R. Kirby, F. King, G. Collet and Y.-Y. Sung
on the carbon fiber scanner, W. Craddock on the gas jet, D. Walz on support systems and R.
Iverson on the beam. And we particularly thank Pisin Chen who proposed the problem of
monitoring the synchrotron radiation.


\clearpage

\clearpage

\clearpage

\clearpage

\clearpage

\clearpage

\clearpage

\clearpage

\clearpage

\clearpage

\end{document}